\documentstyle[preprint,aps]{revtex}
\newcommand{\beq}{\begin{equation}}
\newcommand{\eeq}{\end{equation}}

\begin{document}

\draft

\title{The Reaction $pp\rightarrow pp\eta$ and
the eta-Nucleon and Nucleon-Nucleon Interactions}

\author{M.T. Pe\~na$^{1,2}$, H. Garcilazo $^{1,3}$ and D.O. Riska$^4$}

\address{$^1$ Centro 
de F\'\i sica das Interac\c c\~oes Fundamentais, 1096 Lisboa, Portugal}
\address{$^2$ 
Departamento de F\'\i sica, Instituto Superior T\'{e}cnico, 1096 Lisboa,
Portugal}
\address{$^3$ Escuela 
Superior de F\'\i sica e 
Matem\'aticas, 07738 M\'exico Distrito Federal, Mexico}
\address{$^4$Department of Physics, 00014 University of
Helsinki, Finland}

\maketitle
\begin{abstract}
The $\pi$, $\eta$ and $\rho$ exchange contributions to the cross-section
for the reaction
$pp\rightarrow pp\eta$ near threshold are calculated, with
a phenomenological description of the intermediate S11
(N(1535)) resonance for all exchange mechanisms.    
The final state interaction in the $pp$ system is described by
realistic nucleon-nucleon interaction models.
The sensitivity of the results to the 
phenomenological models for the
$\eta$-nucleon
transition amplitude and the nucleon-nucleon interaction is
explored. The $\eta$ exchange
mechanism is found to play a dominant role. The off-shell behavior of
the $\eta N \to \eta N$ amplitude within
the $\eta-$exchange amplitude leads to a result 
significantly different from that
obtained with the constant scattering length approximation. 
The two-nucleon amplitudes that are associated with the short range
components of the nucleon-nucleon interaction
contribute significantly to the cross section. 
\end{abstract}

\section{Introduction}

The experimental program at electron and hadron beam
facilities as MAMI, GRAAL, COSY, CELSIUS and SATURNE
has brought considerable 
interest for a better theoretical understanding of the
$\eta$-nucleon interaction and 
the reaction mechanisms for $\eta$ meson production.
The recent
experimental results on the total cross section and the angular
distributions for the reaction $pp\rightarrow pp\eta$ very near
threshold 
\cite{Bergdolt,Calen1,Calen2,FZJ}
are very instructive in this regard, especially as
the angular independence of the cross
section in the very near threshold region reveals that the $\eta$-meson 
is produced by a pure $S$-wave mechanism. 

A considerable number of studies of
the reaction mechanisms for $S$-wave $\eta$-production in $pp$
collisions have been carried out. In most of these  
pion-exchange between the protons followed by 
excitation of an
intermediate virtual N(1535) resonance (Fig.1) has
been invoked to describe the reaction mechanism and
the empirical cross
section \cite{Wilkin,Moalem,Meissner}. We here consider 
several additional
contributions to the production amplitude 
besides those hitherto considered. The set of reaction 
amplitudes considered here are illustrated by
the diagrams in Fig.2. These represent the one-nucleon
term (Fig. 2a),
the $\eta$- (Fig. 2b) and $\pi$- (Fig. 2c) 
rescattering terms, and finally the 
amplitudes implied by the 
short-range two-nucleon amplitudes 
which have their origin in the small components of the
nucleon spinors (Fig. 2d).
The "blobs" in Figs.2b and 2c 
represent the full scattering series for the meson-nucleon
transition amplitudes.

The present study addresses the following 4 main
issues: 
 
1) The sensitivity of the calculated cross-section to the 
off-energy-shell $\eta$N scattering
transition matrix. This is motivated by the example of the $\eta$d 
system, for which
the $\eta$d scattering length and the correlated existence and 
position of a quasibound
$\eta$NN state depend sensitively on 
the off-shell behavior of the $\eta$N
amplitude.

2) The contributions to the cross-section of the 
non-resonant two-nucleon
amplitudes that are implied by the short range
components of the two-nucleon interactions. In
the case of the analogous pion production reaction $pp\rightarrow pp\pi^0$
these contributions, as derived from realistic
nucleon-nucleon interaction models, are much larger than the resonance
contributions \cite{RISKA&ME}.

3) The relative weight of the $\pi, \rho$ and  $\eta$ exchange
mechanisms. Even in 
the case of a weak $\eta$NN coupling strength, the present 
calculation 
confirms the result of
ref. \cite{LEESVARC}, which found
the $\eta$ exchange mechanism to be the dominant one. As for the 
magnitude of the
$\rho$-meson exchange contribution, which is somewhat uncertain,
here it is found to be 
non-negligible, if the quark model is used to determine the $\rho
NN(1535)$ transition coupling.

4) The dependence of the calculated cross section 
on the
short range part of the NN interaction. This
is expected to be large for processes
as the present one, for which the
effective momentum transfer between the nucleons is large.

A comment on the  uncertainty in the magnitude of the 
$\eta$ rescattering process is relevant here: this uncertainty
by itself
motivates the present calculation, which attempts to
narrow it by taking advantage of the new experimental information on
the $pp \rightarrow pp \eta$ reaction.
The ambiguity is due to the uncertain value of the $\eta pp$ coupling
constant, $g_{\eta NN}$, which is needed for 
the $\eta$ production vertex in the
process 
illustrated Fig. 2b. The realistic phenomenological NN interaction models
are not sufficiently selective of its value, which they only
constrain into the wide range between
$g _{\eta NN}^2/ 4\pi = 0.25$
and $g _{\eta NN}^2/ 4\pi = 7$, set by the
Nijmegen \cite{NIJM} and one of the BONN \cite{Mach}
potential models respectively. Presumably, only extensions
of these models for energies above the meson production threshold
will show more dependence on the $\eta$NN coupling.
More ambitious
attempts to calculate this coupling constant by
QCD sum rule methods are so far even less
constraining \cite{OKA}.

The uncertainties that arise from the meson exchange description 
that is illustrated in Fig.1 mainly derive from the lack of 
knowledge of
the heavy-meson-nucleon-N(1535) transition couplings,
which cannot be determined
experimentally, as this resonance lies below the threshold for
decay into vector mesons and nucleons. In contrast,
the sub-process  
scattering amplitudes indicated by the blobs in Fig. 2b and 2c have 
been empirically determined in \cite{BAT1,BAT2,GREEN}.

The decomposition into
partial waves $\tau_\ell^j(s)$ 
of the complete unitary series of the 
complete unitary  
$\eta N \to \eta N$ transition amplitude, along with the
$\pi N \to \eta N$ transition amplitude, was
derived
recently by Batini\'c {\it et al.} \cite{BAT1,BAT2}, who made a
fit to all the
available data on the reactions $\pi N \to \pi N$ and $\pi N \to \eta N$.
In ref. \cite{GREEN} fits were made along similar lines to the
empirical data for the reactions $\pi N \to \pi N$,
$\pi N \to \eta N$ $\gamma N \to \pi N$ and 
$\gamma N \to \eta N$
in the energy range of about 100 MeV below and above the $\eta$ production
threshold. Another example of such work, but restricted to the
$\pi N \to \eta N$ amplitude, is given in ref.\cite{FEUST}. 
The model of refs. \cite{BAT1,BAT2} has 
subsequently been tested in a calculation of the 
cross section for the
reactions
$np \rightarrow d\eta$ \cite{TERESA} and 
$\pi d \rightarrow \eta NN$ \cite{GAR1}.
The successful description of the differential cross-section for the
latter reaction, as measured at several energies
\cite{AGS}, shows the quality of the off-energy-shell features
of this model.
This justifies the use of it for the "blob" in Fig. 2b.
For comparison, we used also
the more recent models of the second
work in ref. \cite{GREEN}.

The uncertainty range associated with the 
contributions that are associated with the short-range components
of the nucleon-nucleon interaction is determined by the uncertainty
range of the latter. That in the end turns out to 
be rather modest, however, as it is
constrained
by the requirement on the interaction models of a good description of 
$pp$ scattering data. These uncertainties are therefore far less 
decisive than the 
uncertainty in the two-nucleon contributions that are
associated with the excitation of intermediate N(1535) resonances
by  heavy-meson
exchange between the nucleons (Fig. 1). The uncertainty in these
amplitudes is due,
besides to the uncertainty in the $\eta NN$ coupling
constant, to the very uncertain values of the heavy-meson-N(1535)-N
couplings, as mentioned above. Thus e.g. the
suggestion that $\rho$-meson exchange excitation of the intermediate
N(1535) is completely model dependent \cite{Wilkin2}.

This paper is divided into 5 sections. In section 2 we describe the
amplitudes that are assumed to contribute 
significantly to 
the cross-section. In subsection 2.a the single nucleon amplitude
is described. In subsection 2.b a description of the 
$\pi$ and $\eta$ exchange contributions follows.
Subsection 2.c contains a
description of the short-range exchange amplitudes that are associated
with the nucleon-nucleon interaction. Subsection 2.d describes the
$\rho$-exchange matrix elements.
In section 3 details on the calculation are given.
The numerical results for the cross section are given in section 4.
Finally, section 5 contains a summarizing discussion.

\section{The $\eta-$ production mechanisms}

\subsection{The single nucleon amplitude}

The phenomenologically satisfactory Gell-Mann-Oakes-Renner relations
\cite{GOR} suggest that the octet of light pseudoscalar
mesons, {$\pi,K,\eta$},
are the Goldstone bosons of the broken approximate chiral
symmetry of QCD. Chiral symmetry combined with $SU(3)_F$ symmetry,
then requires that
the coupling of $\eta-$mesons to hadrons 
take the form

\begin{equation}
{\cal L} = -{1\over f_\eta}\partial_\mu\eta A_\mu^8,
\end{equation}
where $A_\mu^8$ is the eight component of the axial vector
current density of the hadron and $f_\eta$ is the $\eta-$meson
decay constant. The empirical value of $f_\eta$ is very close to 
that of
the pion decay constant ($\sim$93 MeV). Production of $\eta$
mesons in $S-$waves in nuclear reactions is then governed by the
axial charge operator $A_0^8$. 

The i=1,2 components of the axial
charge operator are known to have significant two-nucleon
components, which are implied by the nucleon-nucleon interaction
\cite{Mariana,Towner}. These give a large contribution to the
cross section for the reaction $pp\rightarrow pp\pi^0$ near
threshold \cite{Lee}. As the difference between $A_0^\pm$ and
$A_0^8$ only is due to the difference between $\lambda_\pm$ and
$\lambda_8$, it follows from (1) and the Goldberger-Treiman
relation that the pseudovector $\eta NN$ coupling constant
should take the value
\begin{equation}
f_{\eta NN}={g_A\over 2\sqrt{3}}{m_\eta\over f_\eta}\simeq 2.11.
\end{equation}
Here $g_A$ is the axial coupling constant of the nucleon ($g_A$=1.24).
For the corresponding pseudoscalar coupling constant
$g_{\eta NN}=2 (m_p/m_\eta)f_{\eta NN}$ one then obtains the
value 7.23, which is only slightly larger than the value
6.14, which is employed in the Bonn B model for the
nucleon-nucleon interaction \cite{Mach}.
As shown below such a large value for the $\eta NN$ coupling
constant would not allow a satisfactory description of the
experimental cross section for $\eta-$meson
production in $pp$ collisions. Only if the coupling
constant is reduced to about a third of this value, which
is the value that has been
used in phenomenological descriptions of
eta photoproduction \cite{Tiator}, does a satisfactory
description become possible. This indicates that
large $SU(3)_F$ breaking corrections play a significant role in the
chiral Lagrangian (1) in the case of the heavier 
members of the pseudoscalar octet \cite{OKA}. Indeed, if in (2) $m_\eta$
were replaced by the pion mass, as strict $SU(3)$ symmetry would
require, the desired smaller value for $g_{\eta NN}$ would be obtained.

From (1) it follows that the amplitude for 
production of an $S-wave$ $\eta-$meson 
from a single (off-shell)  nucleon may be written as

\begin{equation}
T^{imp}=if_{\eta NN}{\omega_\eta\over m_\eta}{\vec\sigma\cdot
(\vec p\,'_1+\vec p_1)\over 2 m_p} \, \delta(\vec p\,'_2-\vec p_2) +
(1\leftrightarrow 2).
\end{equation}
Here $m_\eta$ and $m_p$ are the masses of the $\eta-$meson
and the proton respectively, $\omega_\eta$ is the energy
of the $\eta-$meson, 
and $\vec p_i$ and $\vec p\,'_i$ ($i=1,2$) are the initial
and final nucleon momenta.

\subsection{The  $\pi N \to \eta N$ and $\eta N \to \eta N$ transitions
and the rescattering amplitudes}

The explicit form of the 
amplitude for a transition
$a N \to b N$, where $a$ and $b$ represent the initial
and final mesons, 
may then be expressed in terms of 
Lorentz invariant amplitudes $A$ and $B$ as
\begin{eqnarray}
t_{a N \to b N}^{\pm} = -A^{\pm}(s,t,u) + 
\frac{1}{2} (q\llap{/}_a + q\llap{/}_b) B^{\pm}(s,t,u),
\end{eqnarray}
where $q_a$ and $q_b$ are 
the 4-momenta of mesons a and b respectively and s,t and u
are the three invariant (Mandelstam) kinematical variables.

For the $\pi N \to \eta N$ and $\eta N \to \eta N$ scattering amplitudes 
we employ the 
variable-mass isobar model described in \cite{GAR1,MOYA}. 
In this model the spin-${1 \over 2}$ and spin-${3 \over 2}$ resonances
are given a mass parameter that is set
equal to $\sqrt{s}$, the invariant mass of the system. The 
meson-nucleon-isobar couplings are given expressions
appropriate for scattering
in states with the orbital angular momentum 
$\ell_{\pm} = j \pm {1 \over 2}$. 

The 
explicit form of the transition amplitude becomes then

\begin{eqnarray}
\lefteqn{t_{a N \to b N} = f_0^{1/2}(s)(P\llap{/} + \sqrt{s}\,)
+ f_1^{1/2}(s)(P\llap{/} - \sqrt{s}\,)} & &\nonumber \\
& & + f_1^{3/2}(s)(P\llap{/} + \sqrt{s}\,)[-q_a\cdot q_b+{q\llap{/}_a 
q\llap{/}_b \over 3} + {2P\cdot q_a P\cdot q_b \over 3s} 
- {P\cdot q_a q\llap{/}_b - q\llap{/}_a P\cdot q_b
\over 3\sqrt{s}}] \nonumber \\
& & + f_2^{3/2}(s)(P\llap{/} - \sqrt{s}\,)[-q_a\cdot q_b+{q\llap{/}_a 
q\llap{/}_b \over 3} + {2P\cdot q_a P\cdot q_b \over 3s} 
+ {P\cdot q_a q\llap{/}_b - q\llap{/}_a P\cdot q_b
\over 3\sqrt{s}}],\nonumber \\
\label{amp}
\end{eqnarray}
\noindent
where $ s = P^2$, and $P$ is the total four-momentum of the 
meson-nucleon subsystem. 

As shown in refs. \cite{GAR1} and \cite{MOYA},
the matrix elements
$\bar u_2(\vec p_2^{\,\prime}) t_{a N \to b N} u_2(\vec p_2)$ in the 
two-body c.m. frame are such that
the functions $f_\ell^j(s)$
are directly related to the phenomenological 
partial-wave
elastic amplitudes in the case $a=b$ and to the partial-wave 
transition amplitudes  
when $a \ne b$. Note that
Eq. (\ref{amp}) is generally valid 
also when any of the four particles 
in the $\pi N \to \eta N$ amplitude are off-mass-shell. 

The $\eta N \to \eta N$ and  
$\pi N \to \eta N$ partial-wave transition amplitudes 
for the $S_{11}$, $P_{11}$, $P_{13}$, and $D_{13}$
channels  were taken from
Batini\'c {\it et al.} \cite{BAT1,BAT2}.
To study the sensitivity of the results to this input we also
included the $\eta N \to \eta$N transition amplitudes given
in ref.\cite{GREEN}.
The $\pi N \to \pi N$ partial-wave amplitudes 
for the $S_{11}$, $S_{31}$, $P_{11}$, $P_{31}$,
$P_{13}$, $P_{33}$, $D_{13}$, and $D_{33}$
channels were taken from
ref.\cite{ARND}.
For the $\eta$ and $\pi$ exchange processes the meson
amplitudes were restricted
to the dominant $S$ waves. 

For the S-wave meson production amplitudes 
in the elastic case a=b one has (see e.g. ref. \cite{WALECKA})
\begin{equation}
[A(s) - (\sqrt s-m_N) B(s)]= \frac{16 \pi s^2}{(\sqrt s  + M)^2 - m_{\eta}^2}
f_0 (\sqrt s), \label{e:kineta}
\end{equation}
with
\begin{equation}
A(s)= -B(s) (m_N+ \sqrt s).
\end{equation}

\noindent

\noindent
At the physical 
threshold, $\sqrt s = m_N+m_{\eta}$, the $\eta N \to \eta N$ strength to
be included in the construction of the $\eta$
exchange amplitude that corresponds to the diagram 2.b 
becomes proportional to the
scattering length $a_0 = f_0(m_N+m_{\eta})$:

\begin{equation}
A\left( (m_N + m_{\eta})^2 \right) - m_{\eta} B
\left( (m_N+m_{\eta})^2 \right)=
4\pi \left(
1+\frac{m_{\eta}}{m_N} \right) a_0.
\end{equation}

As in the present model the loop integration 
for the final-state interaction runs over the available
energy for the  $\eta$N isobar, 
it is natural to define an energy dependent strength as 
\begin{equation}
\lambda_1^{\eta}(s) =\frac{m_{\eta}}{2} [A(s) - (\sqrt s-m_N) B(s)],
\label{e:strength}
\end{equation}
to be inserted into the expression for the 
rescattering diagram. The latter then becomes

\begin{equation}
T^{\eta}= i{g_{\eta NN}\over {2 m_N}}{8\pi\lambda_1^{\eta}(s)\over 
m_{\eta}}
{\vec\sigma^2\cdot \vec k\over m_\eta^2+\vec k^2-\omega_k^2}
f(\vec k^2)+(1\leftrightarrow 2).\label{e:etaex}
\end{equation}
Here $(\vec k,\omega_k)$ is the 4-momentum of the
exchanged $\eta$ meson emitted from a nucleons of initial and 
final three-momentum $p_i$ and $p'_i$
($\vec k = {\vec p}_2 -{\vec p}'_2$ and $\omega_k=
\frac{{\vec k}_2 
\cdot ( {\vec p}_2 +{\vec p}'_2)}{2m_N}$). In the calculations,
the $\eta$-nucleon
invariant mass $s$ is obtained by subtracting the
energy of the spectator nucleon from the total energy of
the NN$\eta$ system. 
The phenomenological form factor
$f$ that dampens out high values of the exchanged momentum is
parametrized as in the Bonn B potential model \cite{Mach}:
\begin{equation}
f(\vec k^2) = \left(\frac{\Lambda_\eta^2 - m_\eta^2}
{\Lambda_\eta^2 + \vec k^2} \right) \, .\label{e:ff}
\end{equation}
Here the $\eta$ cut-off mass is $\Lambda_\eta = 1.5$ GeV.

Analogously, the $\pi$ exchange diagram takes the form 

\begin{equation}
T^{\pi}= i{g_{\pi NN}\over {2 m_N}}{8\pi\lambda_1^{\pi}(s)\over 
m_{\pi}}
{\vec\sigma^2\cdot \vec k\over m_\pi^2+\vec k^2-\omega_k^2}
f(\vec k^2)+(1\leftrightarrow 2),\label{e:piex}
\end{equation}
where $\lambda^{\pi}_1$ is given by an equation analogous to Eq.
\ref{e:strength},
but in which the kinematic factors
of Eq. \ref{e:kineta} are defined by
\begin{equation}
A(s)-(\sqrt s-m_N) B(s) = 4\pi \frac{2 \sqrt s}
{\sqrt(E_1+m_N) \sqrt(E_2+m_N)} f_0 (\sqrt s),
\end{equation}
with the nucleon energies given by
\begin{equation}
E_1=\frac{ s + m_N^2 -m_{\pi}}{2 W},
\end{equation}
\begin{equation}
E_2=\frac{ s + m_N^2 -m_{\eta}}{2 W}.
\end{equation}

The relative sign between $\lambda^{\eta}$ and $\lambda^{\pi}$ is not
fixed by unitarity \cite{LEESVARC}. 
The $SU(3)$ flavor symmetric quark model gives a negative sign
for it \cite{ARIMA}.
As shown below in Section 3 the influence of this sign on the total 
cross-section
of the $pp \rightarrow \eta pp$ is not significant because
of the small effect of
the pion-rescattering diagram.

\vspace{0.5cm}

\subsection{Short-range exchange contributions}

The contributions to the axial exchange charge operator 
that is associated with the
short-range components of the nucleon-nucleon interaction 
have been derived in
refs. \cite{Mariana,Lee} These contributions are 
derived as the nonrelativistic limit of
the nonsingular part of the axial current 5-point function with external leg
couplings. In the case of non-derivative couplings such terms only arise from
the negative energy poles in the nucleon propagator and are 
therefore commonly
illustrated by the nucleon-antinucleon ``pair currents" Feynman diagrams in
Fig.2d. The ultimate 
reason for the appearance of these terms is the Poincar\'e
invariance of the 5-point functions \cite{COESTER}.

Because of the derivative 
pseudovector coupling for both the $\pi$ and the $\eta$, no
such two-nucleon operators 
contribute to the $\eta$ production operator in the
reaction $pp \to pp\eta$, which would be directly 
associated with the $\pi$- and
$\eta$- exchange components of the nucleon-nucleon interaction. The isospin
independent effective scalar and vector exchange components of the
nucleon-nucleon interaction do 
however give rise to two-nucleon meson production
operators. A key 
point is that these operators are completely determined by the
interaction model, and 
involve no parameters that are not already present in the NN
interaction, besides the overall $\eta$-nucleon coupling.

The isospin independent scalar exchange contribution to the two-nucleon
$\eta$-production 
amplitude may moreover be derived directly, without reference
to the 5-point function, in the following way.
Consider the isospin independent scalar exchange component of the
nu\-cleon-nucleon interaction, which is proportional
to the Fermi invariant ``$S$". To
second order in $v/c$, this interaction takes the form

\beq
v_S^+ (r) S = v_S^+ (r) \left( 1-{\vec p\,^2 \over m_N^2}\right) 
- {1\over 2m_N^2}
{\partial v_S^+ \over r\partial r} \vec S \cdot \vec L \ ,
\eeq
where $v_S^+ (r)$ is a scalar function. The $\vec p^2/m^2$ term in the
spin-independent part of this interaction may be combined with the kinetic
energy term in the nuclear Hamiltonian, by replacing the nucleon mass by the
effective ``mass operator"

\beq
m^* (r)=m_N \left[ 1+{v_S^+ (r)\over m_N}\right] \ .
\eeq
To first order in $v_S^+ (r)$, the scalar component of the nucleon-nucleon
interaction 
therefore implies the following two-body ``correction" to the single
nucleon $\eta$ production operator:

\beq
T_2(S) ={v_S^+ (r)\over m_N} T_1 (S) + (1\to2) \ .
\eeq
Here the 
spin-operator in $T_1 (S)$ is implicitly assumed to be that of nucleon
1 of the interacting 
pair of nucleons. This operator coincides in form with the
scalar exchange operator derived in Refs. \cite{Mariana,Towner}.
The corresponding momentum space expression is 

\beq
T_2 (S) = -if_{\eta NN} 
{\omega_{\eta} \over m_{\eta}} {v_S^+ (k_2)\over m_N}
\vec \sigma^1 \cdot \vec v_1 + (1\leftrightarrow 2)\ .
\eeq
Here $v_S^+ (k)$ is the Fourier 
transform of the scalar potential $v_S^+ (r)$
and $\vec v_1 = (\vec p_1 + \vec p_1\,')/2m_N$. This operator is completely
determined by the nucleon-nucleon interaction model. Note that because the
volume integral of the scalar exchange interaction is attractive in 
all realistic nucleon-nucleon
potentials, this exchange current contribution implies an enhancement of the
cross section over the value given by the single nucleon pion production
mechanism.

The expression for the isospin-independent 
vector exchange contribution to the axial charge operator
as derived from the 5-point 
function has been given in Refs. \cite{Mariana,Towner}.
This leads to the following amplitude for $\eta-$meson
production:

\beq
T_2(V^+) = -i {f_{\eta NN}\over m_{\eta}} {v_V^+ (k_2)\over m_N} \left( \vec
\sigma^1 
\cdot \vec v_2 + {i\over 2m_N} \vec \sigma^1 \times \vec \sigma^2 \cdot
\vec k_2 \right)+(1\leftrightarrow 2) \ .
\eeq
Here $v_V^+ (k)$ is the isospin independent vector component of the
nucleon-nucleon interaction.

The corresponding contribution to the $\eta-$meson production
amplitude that is associated with the isospin part
of the vector exchange component of the nucleon-nucleon
interaction is 
\beq
T_2(V^-) = -i {f_{\eta NN}\over m_{\eta}} {v_V^- (k_2)\over m_N} \left( \vec
\sigma^1 
\cdot \vec v_2 + {i\over 2m_N} \vec \sigma^1 \times \vec \sigma^2 \cdot
\vec k_2 \right)\vec\tau^1\cdot\vec\tau^2
+(1\leftrightarrow 2)\ .
\eeq
This term is numerically less important than the previous one that
is associated with the isospin independent vector exchange
interaction, because the latter interaction is by far the stronger
in all realistic nucleon-nucleon interaction models.

\subsection{The N(1535) resonance and $\rho-$meson exchange}

Above the contribution from virtual intermediate $N(1535)$ resonances, 
that are excited by $\pi$ exchange between the two protons, was
incorporated into the scattering amplitude factor $\lambda_1^\pi(s)$
(12). Virtual intermediate $N(1520)$ resonances may, however, also be
excited (or de-excited) by $\rho$-meson exchange (Fig. 1). 
It has in fact been suggested in ref.\cite{Wilkin2} that
$\rho-$meson exchange gives a much larger contribution to
the $\eta-$meson production amplitude than $\pi$
exchange. This argument was based on a comparison of
photoproduction amplitudes, which, however, do not make any
allowance for the mass of the $\rho-$meson.

To derive
this $\rho$-meson exchange amplitude, we employ the transition
Lagrangians
\beq
{\cal L}_{\eta NN(1535)}=i{f_{\eta NN^*}^{(1535)}\over
m_\eta}\bar\psi(1535)\gamma_\mu\partial_\mu\eta \psi +h.c.,
\eeq
\beq
{\cal L}_{\rho NN (1535)}=ig_{\rho
NN^*}^{(1535)}\bar\psi(1535)\gamma_5[\gamma_\mu-{m^*+m_N\over
m_\rho^2}\partial_\mu]\vec\tau\cdot \vec \rho_\mu \psi$$
$$+ h.c.\, .
\eeq
Here $m^*$ denotes the mass of the $N(1535)$, $m_\rho$ the mass of the
$\rho$-meson and $m_N$ the nucleon mass.

The expressions for the two $\rho$-meson exchange amplitudes, which
corresponds to the two Feynman diagrams in Fig. 1 are, respectively,
\begin{eqnarray}
T_1(\rho)=i{f_{\eta NN^*}^{(1535)}\over m_\eta}g_{\rho NN}g_{\rho
NN^*}^{(1535)}{\omega_\eta\over 2m_N}{\vec \tau^1\cdot \vec \tau^2\over
m_\rho^2+k_2^2}f(k_2) \nonumber\\ 
{\vec\sigma^1\cdot \vec k_2-i(1+\kappa_\rho)(\vec \sigma^1\times
\vec\sigma^2)\cdot \vec k_2\over m^*-i\Gamma_r/
2-m_N-\omega_\eta}, \nonumber \\
\end{eqnarray}
\noindent 
and
\begin{eqnarray}
T_2(\rho)=i{f_{\eta NN^*}^{(1535)}\over m_\eta}g_{\rho NN}g_{\rho
NN^*}^{(1535)} {\omega_\eta\over 2m_N}{\vec \tau^1\cdot \vec \tau^2
\over m_\rho^2+k_2^2}f(k_2) \nonumber\\
{\vec \sigma^1\cdot \vec k_2-i(1+\kappa_\rho)(\vec \sigma^1\times \vec
\sigma^2)\cdot \vec k_2\over m^*-i\Gamma_r/
2-m_N+\omega_\eta/2}. \nonumber\\
\end{eqnarray}
\noindent
Here $g_{\rho NN}$ is the $\rho NN$ vector coupling constant and
$\kappa_\rho$ is the ratio of the corresponding tensor to 
vector coupling
constants.
The coupling constant
$g_{\rho NN^*}^{(1535)}$ is that  for the $\rho NN(1535)$ transition
coupling and $f_{\eta NN^*}^{(1535)}$ is that
for the $\eta NN(1535)$ transition
coupling. The resonance width is denoted $\Gamma_r$
respectively. 

The $\eta NN^*(1535)$ coupling constant may be determined directly from
the decay width for $N(1535)\rightarrow N\eta$. This gives the value
$f_{\eta NN^*}^{(1535)}=1.7$
\cite{RISKA&ME}. The $\rho NN(1535)$ transition coupling
constant $g_{\rho NN^*}^{(1535)}$ cannot be determined in the same way,
as the $N(1535)$ lies below the threshold for $N\rho$ decay. 

For the determination of the $\rho NN(1535)$ coupling constant we invoke
the static quark model, which relates the resonance transition coupling
constants to the $\rho NN$ coupling constant $g_{\rho NN}$. 
Employment of the
schematic covariant 3-quark model for the baryons of ref.
\cite{COE2}, which reproduces the empirical baryon spectrum up to
$\sim$ 1700 MeV satisfactorily, leads the following relation \cite{BROWN}:
\beq
g_{\rho NN^*}^{(1535)}=-{\sqrt{2}\over 9}{m_\rho^2\over \omega
m_q}e^{-k^2/6\omega^2}g_{\rho NN}.
\eeq
Here $m_q$ is the constituent quark mass and $\omega$ is an oscillator
parameter for the orbital wave functions, the value of which is $\omega$
= 311 MeV. In this expression $\vec k$ is the momentum of the virtual
$\rho$-meson. If it is set to 0, which is motivated by the
fact that the $\rho$-meson is not far off
shell, the relation (26) gives the value $g_{\rho NN^*}^{(1535)}=-0.88
g_{\rho NN}$, with $m_q=$ 340 MeV.

\section
{Cross-section and distortion in the nucleon-nucleon states}

The total cross section 
for the reaction $pp\rightarrow pp\eta$ may be expressed in
the following form 
after integration over the delta function for conservation
of total four-momentum,
if relativistic three-particle phase-space 
coordinates are employed: 
\begin{equation}
\sigma_{Tlab} [\mu b] = 
10^4 (\hbar c)^2 \frac{m_N^2}{4p_0 E_{p_0}}  \frac{m_N^2}{{4\pi}^3} 
\int_0^{q_{max}} q^2 dq \frac{k'}{{\omega}_q W'} 
|< \psi^-_{p_f} |  M (q) | \psi^+_{p_0}> |^2.
\end{equation}
Here the amplitude $M(q)$ is given in units of 
MeV$^{-3}$ and contains the symmetrization factor $\sqrt{2}$
along with the spin average factor $1/4$.
The initial on-shell relative
nucleon-nucleon momentum is  
$p_0=m_N T_{lab}/2$ and the CM energy 
is $W=\sqrt{4m_N^2+2m_N T_{lab}}$.  The maximal $\eta$ momentum is  
\begin{equation}
q_{max}= \frac{1}{2W} \sqrt{(W^2-4m_N^2-m_{\eta})^2 - 
16 m_N^2 m_{\eta}^2}.
\end{equation} 
The final relative nucleon-nucleon three-momentum in the three
particle CM system is 
denoted $p_f$. Upon boosting to the NN rest frame $p_f$ becomes
$k'$ (which does not differ much from $p_f$ for small eta momentum $q$). 
Furthermore
$W'^2=4(m_N^2+k'^2)=W^2 + m_{\eta}^2 + 2 W \omega_q$ is the corresponding
energy squared, and $\omega_q$ is the 
energy of the emitted $\eta$-meson: $\omega_q= \sqrt{q^2+m_{\eta}^2}$.
The matrix elements of the operators derived in section 2, 
\begin{equation}
 M  =  T^{imp}+ T^{\eta} + T^{\pi} + T_{2}(S) + T_{2}(V),
\end{equation}
were calculated in
momentum space, which allows the non-local
terms in the operators, and the energy dependence arising in the
propagators of the exchanged mesons, to be handled without approximation. 
In principle one has to use the
amplitude distorted by the initial and final state interactions
described by the nucleon-nucleon $T$  matrix in the
expression for the cross section:

\begin{equation}
< \psi^-_{p_f} |  M  | \psi^+_{p_0}>=
< \psi^-_{p_f} |  
T^{imp}+ T^{\eta} + T^{\pi} + T_{2}(S) + T_{2}(V)  | \psi^+_{p_0}>,
\end{equation}
with
\begin{equation}
< k|\psi^+_{p_0}> =  \frac{\delta(k- {p_0} )}
{k^2}  +  \frac{< k | T(E_{p_0}) |{p_0} >}{E_{p_0} -E_k +i\epsilon},
\end{equation}
and 
\begin{equation}
<\psi^-_{p_f}|k> = \frac{\delta(k- {p_f} )}
{k^2}  +  \frac{< {p_f} | T (E_{p_f})|k >}{E_{p_f} -E_k +i\epsilon}.
\end{equation}

We have employed the Bonn and Paris
potentials to calculate the
nucleon-nucleon T-matrix in the final $^1S_0$ nucleon-nucleon
state. The poor
knowledge on the form of the potential 
above the pion production 
threshold ( 300 MeV) makes the issue of the distortion 
in the initial $^3P_0$
state is more 
problematic ( the threshold for $\eta$ production is 1258 MeV).
Instead of extrapolating the potentials mentioned above
to that region, in the absence
of a more realistic extension, for the initial state we used the model
of  ref. \cite{LEEMAT}, as done 
in ref. \cite{LEESVARC}. This reference
gives the ratio of  distortion 
in final and initial states to distortion in  final state only,
as defined by
state distortion factors,
\begin{equation}
f_{ISI}= \frac{ \int_0^{q_{max}} q^2 dq \frac{k'}{{\omega}_q W'} 
|< \psi^-_{p_f} |  M  | \psi^+_{p_0}> |^2} 
{ \int_0^{q_{max}} q^2 dq \frac{k'}{{\omega}_q W'} 
|< \psi^-_{p_f} |  M  | \phi^+_{p_0}> |^2},
\label{eq:dis}
\end{equation}
to be 0.3, a value that will be used here.
The function $\phi^+_{p_0}$ in equation (\ref{eq:dis}) denotes a
plane-wave for the
initial state.
The fact that the reduction factor is constant can be explained by the flat
behavior of the NN T-matrix at high energies \cite{HanhartNa}.
This effect by 
the initial state interaction cannot be discarded if one wants
to extract 
information on the poorly known value of the $g^2_{\eta NN}/4\pi$
coupling constant.

\section
{Numerical Results}

In Fig. 3 the calculated cross section 
for the
$pp \rightarrow pp\eta$ reaction as  
obtained from the 
$\eta$ exchange amplitude, that is represented by the
diagram in Fig.2 b, is shown.
The empirical values in  this and in the following
figures are those given in ref. \cite{Calen1,Calen2}.
The solid curve in Fig 3
gives the result that is obtained with the off-shell 
$\eta N \to \eta N$ amplitude
with energy dependence. The dash-dot-dotted curve gives the
corresponding result obtained in turn with an energy independent
amplitude, of constant strength fixed at the physical threshold 
by the scattering length. The comparison in Fig. 3
shows that the scattering length approximation 
gives rise to an overestimate of a factor of 6 of the 
calculated cross section. 
This striking overestimate indicates that the choice in ref.
\cite{KAISER2}, of using the real part of this scattering 
length to determine
the interaction strength is unrealistic.
 
The $\eta$-exchange result was 
calculated with the value $g^2_{\eta NN}/4 \pi = 0.4$
for the $\eta-$nucleon coupling constant. This
is the value that was obtained by a fit to 
experimental $\eta$ photoproduction 
data in ref.\cite{Tiator}. By scaling up
the results to correspond to the one order of magnitude 
larger coupling constant value that is consistent
with the Bonn potential we find complete agreement with the calculation
of ref. \cite{LEESVARC}, where the same 
phenomenogical $\eta N \to \eta N$
amplitude was used.

Fig. 3 also illustrates that there is a positive interference between the
one-nucleon term and the $\eta$-exchange diagrams as shown by
the short-dashed curve. Once the interference term is taken into
account, the emerging result is close to the result of 
the $\eta$-exchange diagram alone, if  calculated with
a constant strength fixed by the scattering length (dash-dotted line).

In Fig. 4 we show the effect of including the contribution of
the pion exchange mechanism, when it is added to the sum of
the contributions of single nucleon and $\eta-$rescattering 
diagram (dot-dashed line). The $\pi$ exchange contribution
amounts to a $\sim$  15$\%$ increase, if the relative 
sign between the $\pi$ and $\eta$ exchange amplitudes
is taken to be positive.
In Fig. 4 the short-dashed curve has the same meaning as in Fig. 3.
The full line represents the result of 
including in addition the $\rho$-exchange
contribution.

In Fig. 5 we show the effect on the calculated cross section
when the contributions of the short range exchange mechanisms 
that are associated with the nucleon-nucleon interaction
model are added to the previously considered contributions. 
In Fig. 5
the short-dashed line refers to the cross section that
corresponds to
the impulse + rescattering diagrams as in the previous figures,
and the short-dot-dashed line to the 
impulse + rescattering + short-range amplitudes.  The addition
to the impulse and rescattering diagrams of
the short-range amplitudes that are associated
with the isospin independent scalar ("$\sigma$") and
vector ("$\omega$") exchange components of the 
nucleon-nucleon interaction
implies a cross section
that is 9 times larger than before. 
The $\pi$ exchange diagram,  when added 
to the 
impulse$+$rescattering$ 
+$short-range 
amplitudes (dotted line) pushes 
the cross-section even higher, although some
cancellations 
which occur between 
the short-range amplitudes and the $\pi$-exchange reduce
the effect of the last one from 15$\%$, as in Fig. 4, to only 5$\%$. 
The short-range mechanisms also interplay with the $\rho$-exchange
contribution, lessening
its net effect, which nevertheless remains non-negligible (solid line).
If, however, the ratio between the $\rho$NN and the
$\rho$NN(1535) coupling constants is taken 
the same as the ratio of the $\pi$NN and the
$\pi$NN(1535) coupling constants ($\approx$0.18)
the $\rho-$exchange contribution becomes negligible
so that the final result cannot be distinguished from the dotted line.
We consider the results obtained for the $\rho$-exchange with
a quark model based $\rho$NN(1535) coupling constant 
to represent an upper bound for the $\rho$ contribution.
The calculated results shown in Figures 3-5 were obtained
by using the Bonn B potential \cite{Mach} 
along
with the $\eta$N scattering model of reference \cite{BAT1}.

To show the sensitivity of the calculated cross section
values to the nucleon-nucleon potential model, we in
Fig. 6 compare the effects of two different NN interactions
(Paris \cite{Lacombe}: dashed lines and 
Bonn B \cite{Mach}: solid lines)
 to
describe the final state, displaying the uncertainty band originated by the
different short-range behavior of the NN interaction. 
In ref. \cite{GREEN} five different models for
the $\eta N \to \eta N$
transition were given. 
In Fig. 6 the results are shown for the Bonn B and the Paris NN
potentials when model A of the second reference in \cite{GREEN}
is used to describe the $\eta N-\eta N$ transition as well as 
when the model
of reference \cite{BAT1} is used. As to
the other models for the amplitude considered in
ref.
\cite{GREEN}, we have verified that 
models B and C fail to describe the $pp
\rightarrow
pp \eta$ process. Coincidentally or not, these models correspond to
a larger data set
and give a poorer overall fit to the meson-nucleon 
amplitudes considered in that
work. It is worth mentioning that according to \cite{GREEN} a
worse description for the $\gamma N
\rightarrow \pi N$ transition is given by models B and C.
As for model D
of the same reference, we have checked that the corresponding 
results for the $pp \rightarrow pp
\eta$ reaction are close to those obtained 
with model A, and lie in the middle
of the two 
uncertainty regions in Fig. 6, defined by the two border lines relative
to the same NN potential model.

\section{Conclusions}

The present investigation suggests the following main conclusions:

1) The cross section of the reaction $pp \rightarrow pp \eta$ 
in the energy region near the threshold for the $\eta-$production
is very sensitive to the off-energy-shell $\eta$N scattering
transition matrix. 
This sensitivity allows discrimination
between extant phenomenological models for
$\eta$N scattering.
The same sensitivity is observed in the 
$\eta$d interaction, more 
specifically in the magnitude of
the $\eta$d scattering length and the correlated 
existence and position of an
$\eta$NN quasibound state \cite{TERESA}.

2) The non-resonant two-nucleon amplitudes,
the presence of which is implied 
by the short range
components of the two-nucleon interaction, give significant
contributions to the cross section of the reaction. The
situation is in this sense similar to
the case of the corresponding reaction $pp\rightarrow pp\pi^0$.
These contributions, as derived from realistic
nucleon-nucleon interaction models, are moreover much larger than 
the resonance contribution. 

3) The contribution of the $\pi$ exchange mechanism does not 
dominate over that due to $\eta$ exchange - not 
even in 
the case of a very small $\eta$NN coupling.
In ref. \cite{LEESVARC} one
of the same phenomenological models for $\eta N \to \eta N$ scattering
(\cite{BAT1}) that were
considered in this
work was employed in a calculation of the
cross section for $\eta$ production in proton-proton collisions.
When the present calculation was done with the same restricted
input, the results agreed with that in ref. \cite{LEESVARC},
and thus confirm the
finding of that
reference as  to the dominance of the $\eta$-rescattering
amplitude.
The small importance 
of the $\pi$-exchange rescattering diagram relative to the
$\eta$-exchange lessens the importance of the problem of the relative
sign between the $\eta N \to \eta N$ and  $\eta N \to \pi N$ amplitudes,
which is not fixed by unitarity.

4) Only with a value for the $\eta$NN coupling constant 
considerably  smaller than those
used in the earlier realistic nucleon-nucleon potentials
is it possible to obtain a satisfactory description of the
cross section of this reaction with the present model.
Here the same value for the coupling constant was used 
as in the analysis of $\eta$ 
photoproduction data in ref. \cite{Tiator}.

5) The $\rho$-exchange contribution is significant if
the $\rho NN(1535)$ coupling constant obtained in the
constituent quark model is employed.

6) The effect of the nucleon-nucleon interactions in the final 
and in the initial states
are important. In the numerical calculations the
large reduction  from
the initial state 
interaction may appear masked as it may be partially 
compensated by the effect of using a  
constant strength for the $\eta N \to \eta N$ transitions.
The latter procedure was found to represent a poor
approximation, once final and initial state interactions are fully included.

\acknowledgements

D. O. R. thanks Dr. L. Tiator for instructive discussions.
This work was supported in part by 
COFAA-IPN (M\'exico) and
by Funda\c c\~ao para a Ci\^encia e a Tecnologia,
MCT (Portugal) under contracts PRAXIS XXI/BCC/18975/98 and
PRAXIS/P/FIS/10031/1998 and in part by the Academy of Finland
by grants No. 43982 and 44903.

\newpage
\begin{center}
{\bf FIGURES}
\end{center}

Figure 1 - Meson-exchange 
mechanisms that contribute of the reaction $pp \rightarrow
pp\eta$.

Figure 2 - Diagrammatic representation of the amplitudes that
have been considered in this 
calculation. Figs. 2a, 2b, 2c and 2d are the single nucleon, the
$\eta$ rescattering, the $\pi$ rescattering 
and the short-range mechanisms respectively.

Figure 3 - Contribution of the 
$\eta$ exchange diagram  to the cross-section of the
$pp \rightarrow pp\eta$ reaction. The solid line curve
corresponds to a calculation with an off-shell $\eta$N-$\eta$N amplitude
with energy dependence and the dash-dot-dotted one to a calculation 
where the $\eta$N-$\eta$N amplitude strength has its value
proportional to the physical scattering length. 
The short-dashed curve refers to a
calculation where the impulse term is added to an $\eta$-exchange mechanism
built from an off-shell energy dependent $\eta N \to \eta N$ amplitude.
The data points in this and in the following figures are
those given in refs.\cite{Calen1,Calen2}.

Figure 4 - The contribution of 
the $\pi$-exchange diagram  to the cross-section of the
$pp \rightarrow pp\eta$ reaction. The short-dashed curve 
has the same meaning as
in Fig. 3. The dotted-dashed curve includes $\pi$-exchange.
The full line includes the $\rho$-exchange.

Figure 5 - Contribution of the 
short-range mechanisms to the cross-section of the
$pp \rightarrow pp\eta$ reaction. The short-dashed line has the same
meaning as in Figs. 3 and 4, 
and the short-dot-dashed line refers to a calculation with the 
impulse$+$rescattering$+$short-range amplitudes.  The dotted line includes
also $\pi$ exchange and the full line all mechanisms including the
$\rho$-exchange.

Figure 6 - The calculated total cross-section of the
$pp \rightarrow 
pp\eta$ reaction as a function of the energy within the models
considered in
this work.  
The comparison between the calculations with the Paris (dashed lines)
and the Bonn potential (solid lines)
for the final state interaction is made.
Distinction between the $\eta$N amplitude of reference \cite{BAT1} and
the one from model A in the second reference  of \cite{GREEN} is shown.


\begin{references}
\bibitem{Bergdolt} A.M. Bergdolt et al., Phys. Rev. {\bf D48}, R2969 (1993)
\bibitem{Calen1} H. Cal\'{e}n et al., Phys. Lett. {\bf B366}, 39 (1996)
\bibitem{Calen2} H. Cal\'{e}n et al., Phys. Lett. {\bf B458}, 190 (1999)
\bibitem{FZJ} J. Smyrski and P. W\"ustner, Annual Report,
Forschungszentrum J\"ulich, 39 (1998)
\bibitem{Wilkin} C. Wilkin, "Approaches to Threshold Meson Production",
                Proceedings of the {\it International Conference
                Baryons '98}, World Scientific, 505 (1999).
\bibitem{Moalem} E. Gedalin, A. Moalem and L. Razdolskaya, Nucl. Phys.
{\bf A650}, 471 (1999) 
\bibitem{Meissner} 
V. Bernard, N. Kaiser and U.-G. Meissner, Eur. J. Phys. {\bf A4}, 
259 (1999)
\bibitem{RISKA&ME} M. T. Pe\~na, D. O. Riska, and A. Stadler,
                  Phys. Rev. C 60, 045201 (1999)
\bibitem{LEESVARC} 
M. Batini\'c, A. \v Svarc and T.-S. H. Lee, Physica Scripta {\bf 56}, 
               321 (1997)               
\bibitem{BAT1} M. Batini\'c, I. \v Slaus, A. \v Svarc, and B. M. K.  
               Nefkens, Phys. Rev. C {\bf 51}, 2310 (1995)
\bibitem{NIJM} T. Rijken and V. G. J. Stoks, Phys. Rev.
{\bf C59}, 21 (1999)
\bibitem{Mach} 
R. Machleidt, Adv. Nucl. and Part. Phys. {\bf 19}, 189 (1989) 
\bibitem{OKA} H. Kim et al., Nucl. Phys. {\bf A662}, 371 (2000)
\bibitem{BAT2} M. Batini\'c, 
I. Dadi\'c, I. \v Slaus, A. \v Svarc, B. M. K.  
               Nefkens, and T.-S. H. Lee, Physica Scripta {\bf 58}, 15
               (1998). 
\bibitem{GREEN} 
A. M. Green and S. Wycech, Phys. Rev. {\bf C55}, R2167 (1997)
         Phys. Rev. {\bf C60}, 035208 (1999).
\bibitem{FEUST} 
T. Feuster and U. Mosel,  Phys. Rev. {\bf C58}, 457 (1998)
\bibitem{TERESA} H. Garcilazo and M. T. Pe\~na, nucl-th/0002056.
\bibitem{GAR1} H. Garcilazo and M. T. Pe\~na, Phys. Rev. 
               C {\bf 59}, 2389 (1999).
\bibitem{AGS} AGS at Brookhaven National Laboratory: Experiment E890;
              R. Chrien, private communication.
\bibitem{Wilkin2} J.F. Germond and C. Wilkin, Nucl. Phys. {\bf
                A518}, 308 (1990) 
\bibitem{GOR} M. Gell-Mann, R. Oakes and B. Renner, Phys. Rev. {\bf 175},
2195 (1968)
\bibitem{Mariana} 
M. Kirchbach, D. O. Riska and K. Tsushima, Nucl. Phys. {\bf A542},
616 (1992)
\bibitem{Towner} I. Towner, Nucl. Phys. {\bf A542}, 631 (1992)
\bibitem{Lee} T.-S. H. Lee and D. O. Riska, Phys. Rev. Lett. {\bf 70},
2237 (1993)
\bibitem{Tiator} L. Tiator, C. Bennhold and S. S. Kamalov, "The eta
coupling and the $S_{11}$ resonance in eta photoproduction on the
nucleon", Physics with GeV-Particle Beams", H. Machner and
K. Sistemich, World Scientific, 112 (1995).
\bibitem{MOYA} H. Garcilazo and E. Moya de Guerra, Phys. Rev. C {\bf 52}, 
               49 (1995).
\bibitem{ARND} R. A. Arndt, Z. Li, L. D. Roper, and R. L. Workman,    
               Phys. Rev. D {\bf 43}, 2131 (1991).
\bibitem{WALECKA} J. D. Walecka, {\it Theoretical Nuclear and Sub-Nuclear
Physics}, Oxford University Press, New York, (1995). 
\bibitem{ARIMA}M. Arima, K. Shimizu, K.Yazaki, Nucl. Phys. A543,613 (1992).
\bibitem{COESTER} F. Coester and D. O. Riska, Ann. Phys. (N. Y.)
               {\bf 234}, 141 (1994).
\bibitem{COE2} F. Coester and D. O. Riska, Few-Body Systems
{\bf 25} (1998) 29
\bibitem{BROWN} D. O. Riska and G. E. Brown, eprint nucl-th/0005049

\bibitem{LEEMAT} 
T-S. H. Lee and A. Matsuyama, Phys. Rev. {\bf C36}, 1459 (1987)
\bibitem{HanhartNa}C. Hanhart, K. Nakayama, Phys. Lett. B454, 176 (1999).
\bibitem{KAISER2} N. Kaiser, Proceedings of the 7th Conference {\it
Mesons and Light Nuclei '98}, Edited by J. Adam, P Bydzovsky, J. Dobes,
R. Mach, J. Mares, M. Sotona, World Scientific, (1999).
\bibitem{Lacombe} M Lacombe et al., Phys. Rev. {\bf C21}, 861 (1980)


\end{references}
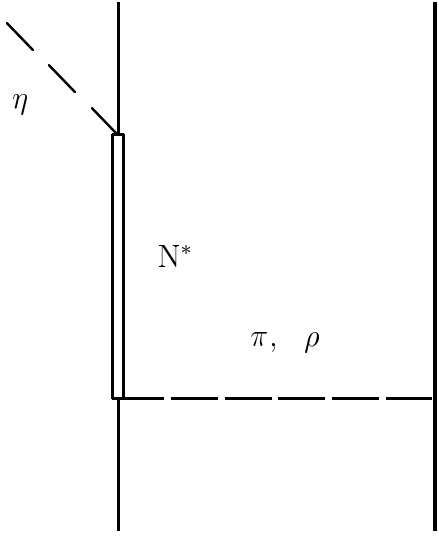
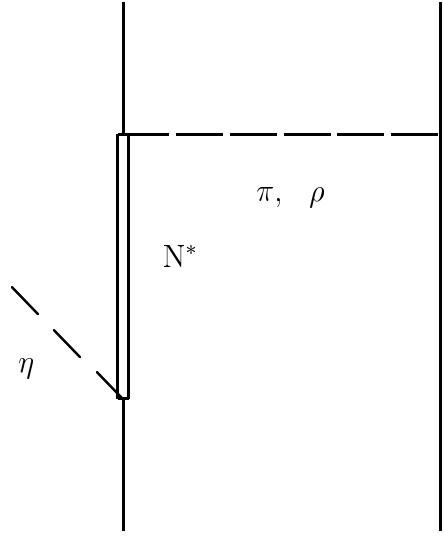
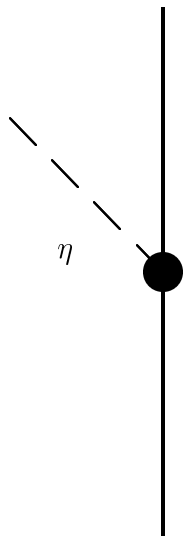
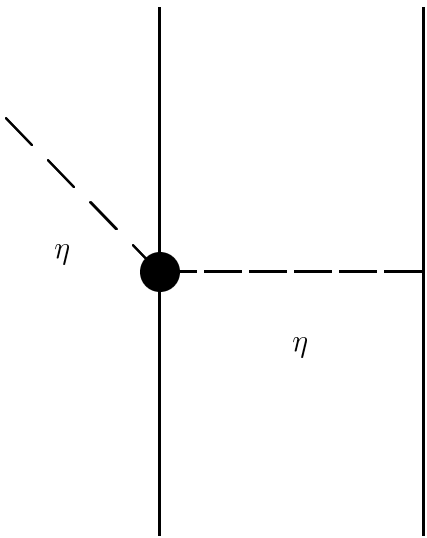
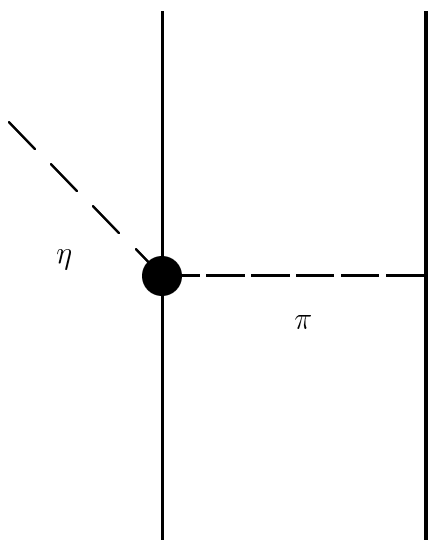
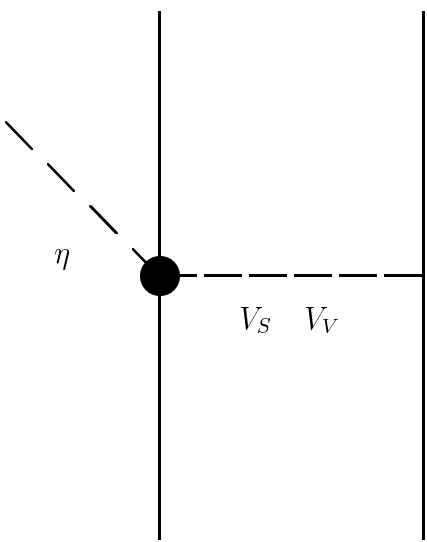
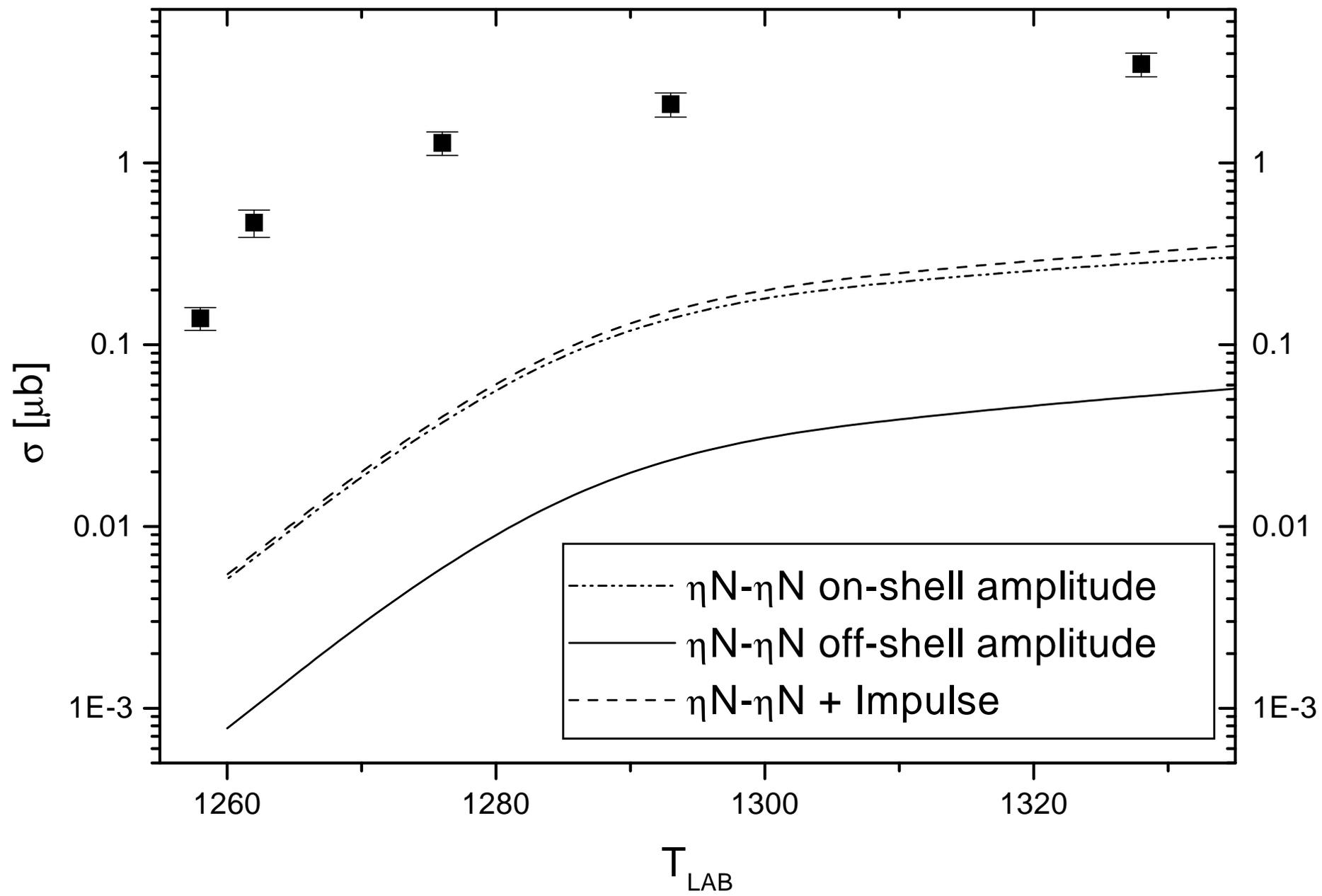
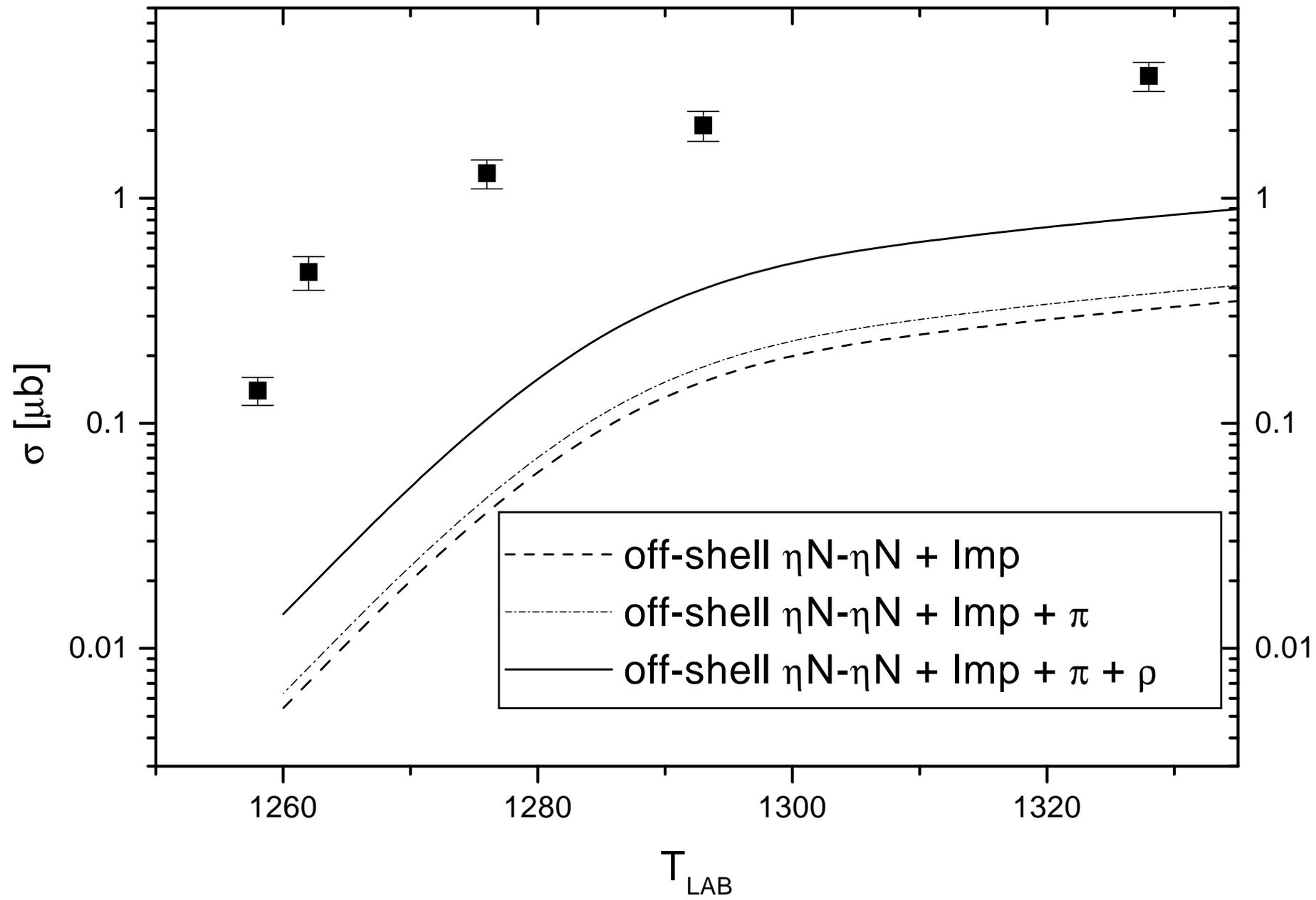
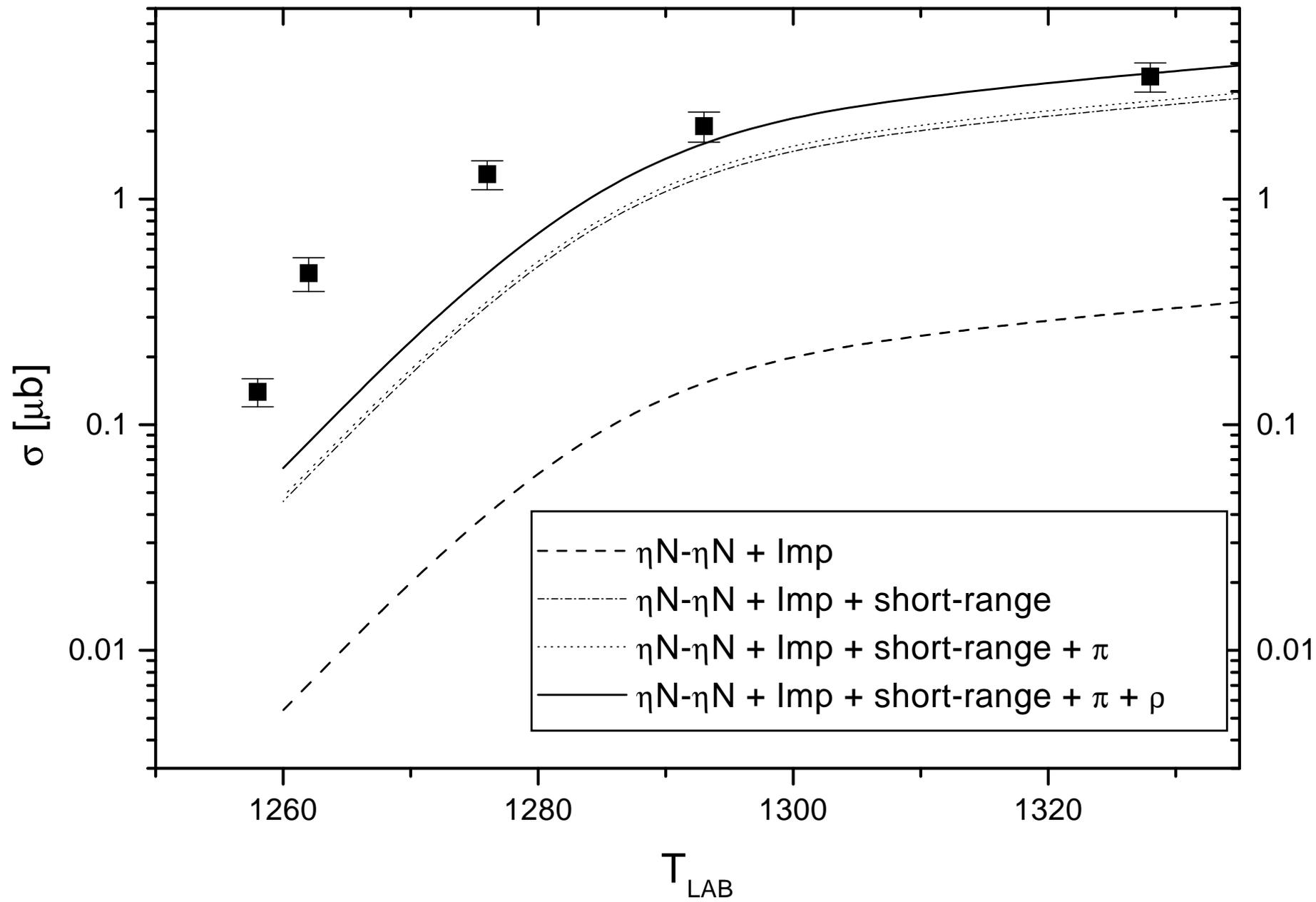
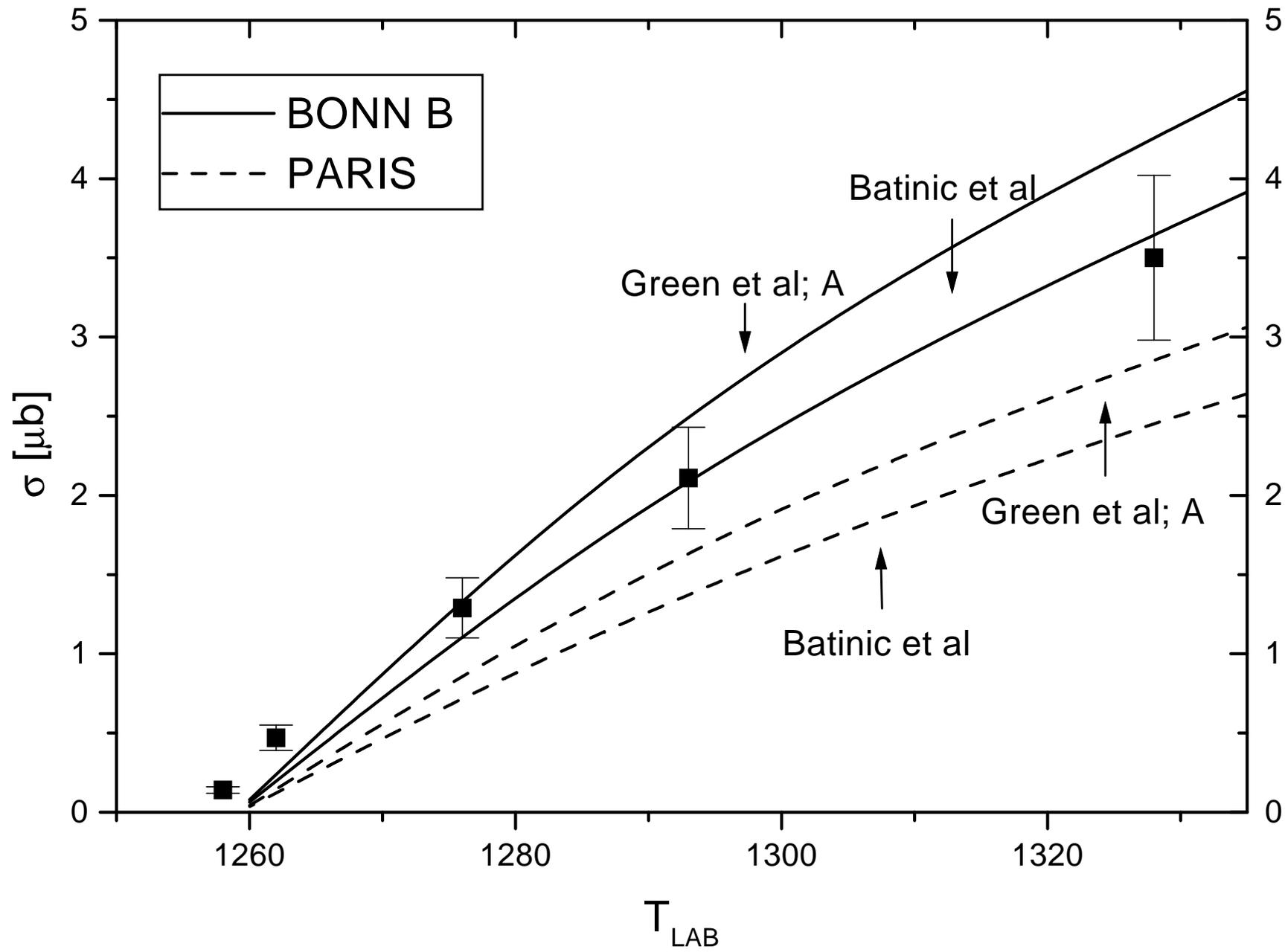
\end{document}